\documentclass{article}

    \PassOptionsToPackage{numbers, compress}{natbib}

 \usepackage[creativeai, final]{neurips_2025}
\usepackage{graphicx}

\usepackage{xcolor}

\usepackage{xcolor}
\usepackage{soul}

\usepackage{tikz}
\usetikzlibrary{
    shapes.geometric, 
    arrows.meta,      
    positioning,      
    backgrounds,      
    fit               
}

\usepackage{booktabs}  
\usepackage{tabularx}  
\definecolor{thoughtcolor}{gray}{0.90}

\usepackage[utf8]{inputenc} 
\usepackage[T1]{fontenc}    
\usepackage{hyperref}       
\usepackage{url}            
\usepackage{booktabs}       
\usepackage{amsfonts}       
\usepackage{nicefrac}       
\usepackage{microtype}      
\usepackage{xcolor}         

\title{Panel-by-Panel Souls: A Performative Workflow for Expressive Faces in AI-Assisted Manga Creation}
\author{
  Qing Zhang \\ The University of Tokyo  \\ \href{mailto:qzkiyoshi@gmail.com}{qzkiyoshi@gmail.com}, \And
  Jing Huang \\ Tokyo University of the Arts \\ \href{mailto:hkoukenj@gmail.com}{hkoukenj@gmail.com}, \And
  Yifei Huang \\ The University of Tokyo \\ \href{mailto:mingyang@kmd.keio.ac.jp}{hyf015@gmail.com}, \And
  Jun Rekimoto \\ The University of Tokyo \\SONY CSL Kyoto\\ \href{mailto:rekimoto@acm.org}{rekimoto@acm.org} \\
}

\begin{document}

\maketitle

\begin{abstract}
Current text-to-image models struggle to render the nuanced facial expressions required for compelling manga narratives, largely due to the ambiguity of language itself. To bridge this gap, we introduce an interactive system built on a novel, dual-hybrid pipeline. The first stage combines landmark-based auto-detection with a manual framing tool for robust, artist-centric face preparation. The second stage maps expressions using the LivePortrait engine, blending intuitive performative input from video for fine-grained control. Our case study analysis suggests that this integrated workflow can streamline the creative process and effectively translate narrative intent into visual expression. This work presents a practical model for human-AI co-creation, offering artists a more direct and intuitive means of ``infusing souls'' into their characters. Our primary contribution is not a new generative model, but a novel, interactive workflow that bridges the gap between artistic intent and AI execution.

\end{abstract}

\section{Introduction}
In manga (Japanese-style comics known for their expressive art and panel-based storytelling) a character's soul is conveyed in the subtle shift of an eye or the slight curve of a lip, a nuance that current text-to-image models consistently fail to capture. While state-of-the-art text-to-image models can generate aesthetically striking manga-style characters, they struggle with rendering the subtle, coordinated facial dynamics essential for sequential storytelling \cite{wu2025diffsensei}. A manga artist's intent behind a ``knowing glance,'' ``a flicker of suspicion,'' or ``a smile tinged with regret'' is often lost in translation. This results in compositionally sound but emotionally vacant panels, or worse, expressions that are inconsistent from one panel to the next, breaking the narrative thread \cite{wu2025diffsensei}. This ``nuance gap'' forces a manga artist into a laborious fine-tuning stage, manually redrawing faces to ensure the story flows correctly.

Existing digital solutions, however, present their own set of challenges for the practicing manga artist. Powerful facial reenactment tools are primarily designed for video production, with interaction models ill-suited for the iterative refinement of static manga panels \cite{luo2025dreamactor, liao2025enhancing}. Alternatively, highly flexible toolkits like ComfyUI require artists to navigate complex, node-based editors. This disjointed process breaks the creative ``flow state,'' which is crucial for a manga artist, who often works under tight deadlines and whose primary tool is a layer-based drawing application like Clip Studio Paint, not a visual programming environment. There is a clear gap for an integrated workflow designed specifically for the narrative needs of a manga artist in terms of human-AI co-creation: one that is intuitive, direct, and allows for the fluid orchestration of expressions.

We address this need with a novel, dual-hybrid pipeline designed to balance generative AI efficiency with artistic control. Our workflow begins with a hybrid preparation stage that combines a landmark-based auto-detector for speed with a manual framing tool for handling complex hairstyles or accessories \cite{yu2025geneava}. This is followed by a hybrid expression mapping stage, where broad emotions are captured through performative video input and then perfected with fine-grained numerical sliders \cite{luo2025dreamactor, yu2025geneava}. This multi-stage, artist-centric approach allows the manga artist to remain focused on character dynamics and emotional storytelling, using the system as a powerful, collaborative partner rather than a rigid tool.

Our primary contribution is a novel system and workflow that serves as a constructive model for human-AI co-creation. In the context of the changes and risks brought by generative AI, our work presents an alternative that leverages, rather than supplants, human creativity \cite{haribonpublishingToolsCompanions}. We offer both a concrete example of human-AI co-creation and a direct approach targeting the fundamental gap between text and nuanced expression. Specifically, this paper introduces a performative pipeline for managing individual character expressions within multi-character or sequential panels. This system contributes a practical tool and a new workflow that empowers artists to infuse their characters with emotion in a more direct and intuitive manner.

\section{Related Work}
\subsection{The Expressive Challenge in Manga: Beyond the Text Prompt}

The advent of powerful diffusion models like Stable Diffusion and DALL-E has enabled the generation of high-quality, manga-style illustrations from simple text prompts \cite{kawar2023imagic}. While adept at capturing overall artistic style, they frequently fail to render the subtle emotional states or micro-expressions with the precision required for sequential storytelling \cite{wu2024portrait3d}. This issue stems from a fundamental cognitive-linguistic gap: our rich mental imagery of a character’s expression is inherently resistant to complete verbal encoding \cite{peper2022general, schiffer2017intention,mathews2013feels,monzel2024affective,nakabayashi2008role}. Researchers have noted this barrier in general AI image creation, where users find it difficult to articulate their desired facial expressions with adequate precision \cite{karras2023generative, melendez2018beyond}.

This challenge is amplified in manga, where storytelling relies on a unique and complex visual lexicon. A trope like \textit{tsundere}, for instance, requires a subtle blend of anger, affection, and embarrassment \cite{abbott2011visual} that defies a simple text description. Similarly, conveying the silent, panel-to-panel shift in a team captain’s expression from steadfast confidence to quiet concern is nearly impossible to orchestrate through iterative text prompts \cite{animationstudiesDanielJohnson}. Relying on text alone often forces the manga artist to accept generic, stereotypical emotions that flatten the narrative depth.

\subsection{The manga artist's Digital Toolkit: Current Workflows and Frictions}\label{sec:current_workflow}

To overcome the limitations of generative models, a manga artist typically reverts to their established digital toolkit, dominated by layer-based software like Clip Studio Paint \cite{clipstudioCLIPSTUDIO}. Within this environment, two primary methods for expression control exist: direct manual drawing and the use of 3D posing models. Manually redrawing each face panel-by-panel offers maximum control but is intensely time-consuming and creates a high risk of introducing subtle inconsistencies in a character's appearance. The alternative, using built-in 3D posing dolls, helps maintain consistency but introduces new frictions. These models are often criticized for their generic appearance and stiffness, requiring significant manual alteration to match a character's unique style and inject life into their expression \cite{Endangered, Aiconomist}.

While advanced visual programming interfaces for diffusion models, such as ComfyUI, offer powerful fine-tuning capabilities via node-based editors, their interaction paradigm is fundamentally disconnected from the manga artist's core drawing process. Requiring a manga artist to switch from a digital canvas to a complex graph editor breaks the creative ``flow state'' crucial for narrative work. These tools, while powerful, are not designed with the workflow of a sequential artist in mind, creating a clear human-AI interaction gap for a more integrated and intuitive solution \cite{Not4Talent}.

\subsection{The Workflow Gap: Integrating Reenactment for Co-Creation}

Facial reenactment methods \cite{rochow2024fsrt, guo2024liveportrait, zhao2025x} offer compelling alternatives, enabling direct ``performative'' input to capture artists' desired expressions. However, these methods introduce their own significant workflow gap. State-of-the-art reenactment models, much like the powerful node-based editors discussed in Section \ref{sec:current_workflow}, are rarely designed for a manga artist's panel-based workflow. They are often standalone tools for video production or exist as components within complex, non-intuitive interfaces that, as previously noted, break the creative ``flow state''.

The true gap is therefore the lack of a system that leverages generative AI for what it excels at—rapidly drafting scenes and characters—while seamlessly integrating an intuitive, performative system that allows the artist to re-take control and ``infuse souls'' into their characters' expressions panel-by-panel. Our work addresses this gap by integrating a high-fidelity reenactment engine into a novel interaction model, designed to bridge the needs of the manga artist with the capabilities of modern AI.

\section{Our Approach: The Panel-by-Panel Pipeline}
Our approach transforms an initial AI-generated manga panel into a narrative illustration through a three-stage pipeline designed for a manga artist's workflow: (1) Automated Face Preparation, (2) Interactive Expression Mapping, and (3) Layered Composition and Refinement. This process is designed to be repeated seamlessly across multiple characters and panels, allowing for the consistent development of emotional storytelling. The central hypothesis of our work is that the hybrid nature of the Interactive Expression Mapping stage is the key to this workflow's effectiveness, a claim we explore through a focused analysis in Section \ref{sec:case_study}.

\textbf{Stage 1: A Hybrid Pipeline for Artist-Centric Face Preparation.} The pipeline begins with a hybrid face preparation stage, designed to balance automated efficiency with artist-centric control. We recognize that while automated face detectors are powerful, the creative scope of manga—with its complex hairstyles, face-obscuring accessories, and varied character orientations—demands a system that ensures the artist always has the final say.

Our system therefore provides two integrated modes: (1) Landmark-Based Auto-Detection: The primary mode utilizes a state-of-the-art face analysis model from the insightface library \cite{deng2019arcface}. Instead of relying on a simple bounding box, this method detects 106 facial landmarks (eyes, brows, nose, mouth, and facial contour). We then algorithmically construct a tight, padded bounding box directly from these landmarks. This technique is highly effective at producing well-composed crops of human-like faces while explicitly excluding non-facial elements like hands and torsos, a critical failure point of general object detectors. (2) Manual Framing Mode: For cases where the automated detection is insufficient or when the artist wishes to make a specific creative choice (e.g., including a particular accessory as part of the `face'), the system provides a seamless manual mode. The artist is presented with an interactive window where they can directly draw a square frame onto the source image. This manual bounding box is then treated identically to one generated by the auto-detector.

This dual-mode pipeline provides the speed of automation for standard cases while preserving the ultimate creative control and robustness of manual intervention, guaranteeing that any character can be prepared for the reenactment stage.

\textbf{Stage 2: Interactive Expression Mapping.} Continuing the hybrid design philosophy from the preparation stage, our Interactive Expression Mapping workflow is also built on a dual-mode interaction model designed to offer both intuitive performance and precise control. Our system leverages the pre-trained LivePortrait \cite{guo2024liveportrait} model for its balance of real-time efficiency and high-fidelity output. We integrate this engine into a novel, hybrid interaction workflow that offers both intuitive performance and precise control.

First, the manga artist provides a performative input, either through a live webcam feed or by uploading a pre-recorded video reference, a common practice for artists seeking to capture specific expressions. Our interface presents this video on an interactive timeline, allowing the artist to scrub through the performance and select the single keyframe that best captures the desired macro-expression, such as the peak of a smile or a specific head tilt.

Recognizing that a single performance frame may lack precision, the system then offers a secondary stage of fine-grained numerical control. After the base expression is mapped, the manga artist can use dedicated sliders to directly manipulate parameters of the eye and lip retargeting modules. This hybrid approach is allowing for adjustments that are difficult to perform, such as correcting a character's gaze independently of their head pose or precisely modifying the lip curvature to match a character's established design. This gives the manga artist the expressiveness of a reference performance combined with the meticulous control needed for detailed line art.

\textbf{Stage 3: Composition and Refinement.}
The final stage of our pipeline focuses on the practical task of re-integrating the expressive face from Stage 2 back into the original manga panel. The process begins with a direct composition method: the 512x512 pixel reenacted face is first resized back to its original dimensions as recorded during Stage 1. It is then pasted onto the source panel at its precise original coordinates.

We acknowledge that this direct composition, while efficient, introduces predictable artifacts along the seams of the pasted region. These may include subtle geometric misalignments where the new face meets the character's neck and hair, or minor shifts in hue and lighting introduced by the reenactment model. A simple paste leaves a visible, jarring edge that requires refinement.


While automated seam-blending \cite{zhang2020deep} could potentially address these artifacts, our work's primary contribution is the novel expression mapping pipeline (Stages 1 and 2). We therefore treat the final integration as a separate, known post-processing challenge. Our workflow intentionally concludes by handing off a composited draft, with its predictable artifacts, to the artist. This approach provides a clean hand-off point, allowing professional manga artists to apply their own expertise and familiar tools such as a smudge or airbrush brush in their preferred software for the final, nuanced polishing.


This design choice reinforces our vision of the system as a powerful assistant rather than a complete replacement. It automates the laborious task of redrawing faces for nuanced expressions—isolating characters and mapping complex expressions—while leaving the final, delicate task of aesthetic integration and polishing firmly in the hands of the artist. This respects their role as the ultimate arbiter of quality and maintains their creative control over the final artwork.

\begin{figure}
    \centering
    \includegraphics[width=\linewidth]{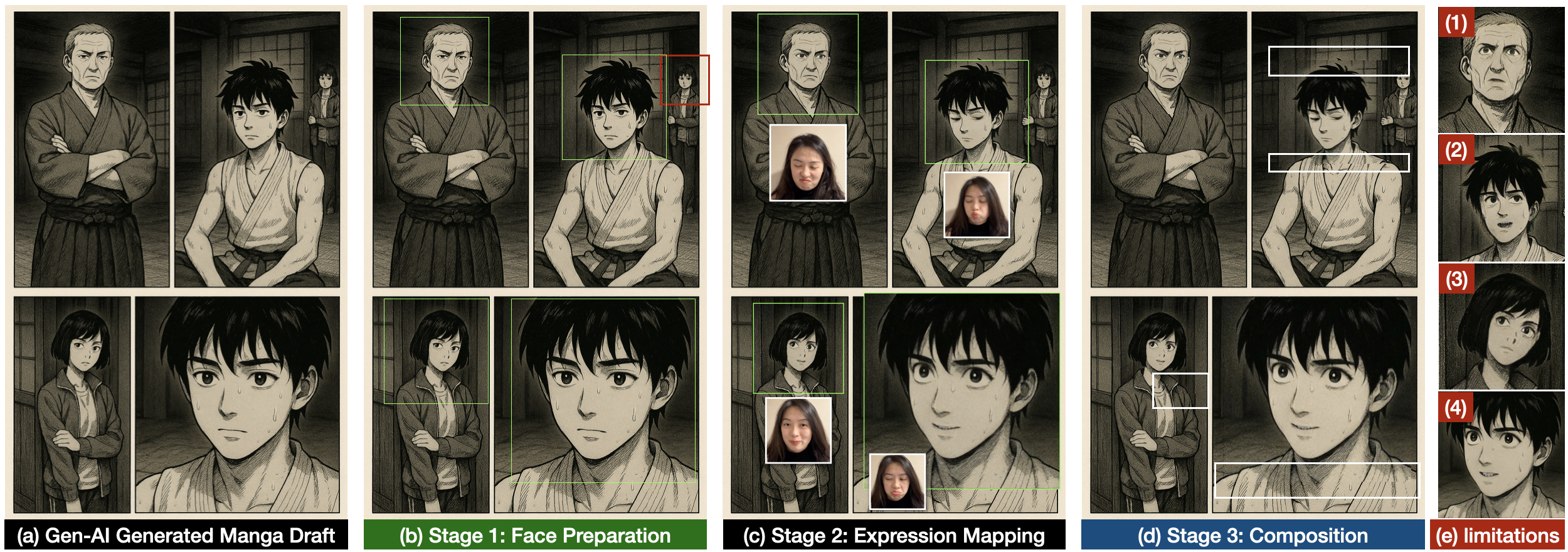}
    \caption{An end-to-end demonstration of our pipeline and an illustration of its current limitations. The process begins with (a) a Gen-AI generated manga draft with neutral character expressions. In (b) Stage 1: Face Preparation, our system successfully identifies and frames the primary faces, though it fails on a small, distant face (highlighted in red), illustrating a limitation of the auto-detector. During (c) Stage 2: Expression Mapping, new expressions are mapped onto the prepared faces from driving inputs (shown inset). In (d) Stage 3: Composition, the modified faces are re-integrated back into the draft, demonstrating a successful transfer of expression. Finally, (e) highlights typical limitations, which are largely inherited from the underlying LivePortrait model.}
    \label{fig:placeholder}
\end{figure}

\section{Analysis of an End-to-End Case Study}\label{sec:case_study}
To evaluate the practical application and effectiveness of our pipeline, we conducted an end-to-end case study on a multi-character, multi-panel manga draft that was initially generated with neutral expressions using a state-of-the-art text-to-image model (DALL·E 3). The goal was to use our system to infuse the characters with new expressions to alter the narrative tone of the scene. The entire process and its outcomes are illustrated in Figure \ref{fig:placeholder}. This case study was conducted as a formative and expert evaluation. It was performed by the authors, who possess formal training in fine arts (visual communication design, painting) and human-computer interaction, allowing for a dual analysis of both the technical pipeline and its alignment with an artist's creative workflow.

Character Prompts:
\textit{"""Character 1 (Older Mentor): Middle-aged man with stern features, short graying hair, wearing a worn hakama and gi. Neutral, composed expression. Standing with arms folded, upright posture, authoritative presence.
Character 2 (Young Trainee): Teenage boy, lean build, messy black hair, wearing a sleeveless gi. Sitting on the floor, legs crossed, looking forward with a blank, neutral expression. Sweat marks his brow, but his face is calm.
Character 3 (Observing Peer): Teenage girl, short bob haircut, wearing a tracksuit top over training clothes. Leaning against the doorframe with one arm loosely folded. Neutral expression, gaze directed toward the seated boy."""}

\textbf{Findings from Stage 1: Face Preparation.} As shown in Figure \ref{fig:placeholder}(b), our landmark-based auto-detector successfully identified and framed the primary male characters. However, this process revealed two key insights. First, our findings suggest that characters with complex hairstyles or accessories often require the manual face extraction mode to achieve an artistically sound composition. This finding justifies our hybrid approach. The auto-detector's failure on the small, distant face (highlighted in red) further underscores the need for a manual override. Second, we found that a slightly generous crop that includes a character's hair works better for the subsequent reenactment stage than a tight crop on only the facial features. This provides more contextual information to the reenactment model, leading to a greater perceived coherency between the face and hairstyle in the final result.

\textbf{Findings from Stage 2: Expression Mapping.} Figure \ref{fig:placeholder}(c) illustrates the core expression mapping workflow. A crucial observation during this stage was the discrepancy between the driving performance and the final reenacted output. The most aesthetically pleasing or narratively correct facial expression on the manga character would often appear a few frames before or after the seemingly best frame of the driving video. This temporal offset highlights a key human AI interface challenge: the user's perception of their own best performance may not always align with the model's best interpretation. This finding underscores the necessity of our interactive timeline slider, which allows the artist to scrub through the results to find the optimal frame, rather than being locked into a single moment from their initial performance.

\textbf{Findings from Stage 3: Composition and Limitations.} The successful re-integration of the new faces is shown in Figure \ref{fig:placeholder}(d). However, the limitations of the underlying reenactment model become apparent here, as detailed in Figure \ref{fig:placeholder}(e). We observed that no matter whether we used relative or absolute motion modes in LivePortrait, a set of typical artifacts still appeared. These persistent artifacts include geometric inconsistencies where the hair and ears remain static during head rotation (e1-e3) and style mismatches where photorealistic features like lips and teeth are introduced into the monochrome aesthetic (e4). This observation reinforces that current reenactment methods are not a perfect one-shot solution.

\section{Discussion}

Our end-to-end case study provided several key insights into the practical challenges of human-AI co-creation for manga art. The analysis confirmed the necessity of our hybrid pipeline, as the automated face detector's failure on distant characters or complex hairstyles highlighted the need for a manual override to handle artistic diversity. More importantly, the expression mapping stage revealed a crucial human-AI co-creation challenge: a temporal offset, where the most aesthetically pleasing or narratively correct reenacted expression did not always align with the perceived best frame of the artist's driving performance.

This temporal discrepancy finding is central to our contribution. It demonstrates that a simple, one-shot ``perform-and-generate'' model is insufficient for this creative task. Instead, it validates our workflow's emphasis on an interactive timeline slider, which allows the artist to scrub through the results and discover the optimal frame. This transforms the interaction from a rigid command into a more fluid, collaborative exploration.

The importance of this human-in-the-loop approach is underscored by existing concerns within the professional manga community. In an expert interview conducted by the authors after the case study, a professional manga artist noted that the current widespread adoption of digital tools (e.g., pen tablets with line smoothing) has already led to a perceived ``stylistic homogenization''—a loss of the unique, individual ``pen stroke'' characteristic of traditional, physical-media artwork.

This existing concern over stylistic homogenization provides a crucial parallel to the problem of expressive homogenization our paper addresses: the ``emotionally vacant'' or generic faces produced by text-to-image models. Our workflow, therefore, can be seen as a counter-movement. Instead of further automating and ``smoothing'' the artist's unique hand, it is designed to capture and translate the artist's personal, performative intent. It serves as a constructive model for human-AI co-creation by re-centering the artist, empowering them to ``infuse souls'' through their own performance while leaving the final aesthetic integration firmly in their control.

Despite its successes, our proof-of-concept also has clear limitations. The most significant is the lack of holistic, 3D-aware understanding, which results in artifacts where a character's head rotates but their hair and ears remain static. Furthermore, the stability of the facial reenactment is sensitive to head pose. Echoing the challenges faced by foundational models \cite{wu2024portrait3d}, we observed that faces turned more than 45 degrees from the camera are still problematic, leading to less reliable expression mapping. Style mismatches can also occur, where photorealistic features are introduced into a stylized artwork, breaking the aesthetic cohesion and suggesting the necessity of a fine-tuned model specifically for this artistic domain. Finally, our current landmark-based auto-detector, while effective for human-like faces, is not designed to handle the full range of non-human characters common in manga creation. We acknowledge that facial reenactment technologies carry dual-use risks, but our work's focus remains on the constructive application of these tools to empower artists.

These limitations provide a clear roadmap for future work. To address the geometric inconsistencies, the next iteration of our system should integrate 3D-aware models, potentially leveraging ControlNet-like mechanisms for pose control \cite{zhang2023adding}. Such an approach would allow artists to manipulate not just the facial muscles but the character's entire head orientation and posture, offering unparalleled freedom for narrative creation. To combat style mismatch, future research could explore fine-tuning the reenactment model on a curated dataset of manga-style art or incorporating style-preserving techniques.

Ultimately, we see our system as a step that could transform the artist's role from a puppeteer of expressions to a true digital sculptor of character performance. By offloading the laborious task of redrawing faces for every subtle emotional shift, our work suggests that the most promising path for creative AI is not one that seeks to replace the artist, but one that provides them with powerful, intuitive, and collaborative tools to bring their visions to life more effectively. A key direction for future work is to conduct a formal user study with professional manga artists, moving beyond our initial author-led evaluation, to quantitatively assess our system's impact on workflow efficiency and qualitative ratings of expressive control. The expert interview conducted for this paper confirmed the relevance of our research problem. The key human-AI co-creation challenges identified in our initial case study—such as the temporal offset in expression mapping and the necessity of the manual framing tool for complex compositions—will serve as the primary hypotheses and design probes for this future formal evaluation with industry professionals.

\section{Conclusion}
This paper introduced a dual-hybrid, performative workflow for AI-assisted manga creation. Our analysis indicates that this artist-centric approach is potentially a viable and effective method for translating narrative intent into visual art. This research contributes a practical pipeline that streamlines a tedious creative task, serving as a constructive model for human-AI co-creation in visual storytelling.

\begin{ack}
We sincerely thank professional manga artist YUNA (Mingxi Zhang) for her invaluable time and expert insights into the professional digital manga workflow. This work was supported by JST Moonshot R\&D Grant JPMJMS2012, JSPS Grant-in-Aid for Early-Career Scientists JP25K21241, JSPS KAKENHI JP25K24384, and JPNP23025 NEDO.
\end{ack}


\end{document}